\documentclass[paper,notoc]{JHEP} 
\usepackage{epsfig}
\usepackage{amsmath}
\usepackage{amssymb}
\usepackage{cite}
\psfull

\def\kk{{\bf k_1}}
\def\kg{{\bf k_2}}
\def\qq{{\bf q}}

\title{Truncated BFKL Series and Hadronic Collisions} 
\author{M.B. Gay Ducati and M.V.T. Machado\\  
  Instituto de F\'{\i}sica, Univ. Federal do Rio Grande do Sul.
 Caixa Postal 15051, 91501-970 Porto Alegre, RS, BRAZIL. \\
  E-mail: \email{gay@if.ufrgs.br}, \email{magnus@if.ufrgs.br}}

\abstract{ We study the contribution of a truncated BFKL Pomeron series to the hadronic
processes showing that a reliable description is obtained using two orders in
perturbation theory. The $pp$($p\bar{p}$) total cross sections are described
with  good agreement, consistent with the unitarity bound. We also calculate
the elastic  scattering amplitude at non zero momentum transfer $t$,
introducing two distinct ans\"atze for the proton impact factor. As a by
product the elastic differential cross section is obtained at small $t$
approximation and compared with the data, describing with good agreement this
observable for both low and high energies values. }


\begin{document}
\section{Introduction}

Several years ago was started the calculation program of the perturbative
contribution to the Balitsky, Fadin, Kuraev and Lipatov (BFKL) Pomeron,
generating the integral equation which determines its behavior
in perturbative QCD \cite{bfkl}. That procedure consists of summing the leading
logarithms on energy (LLA), $\ln(s)$, order by order from perturbation theory,
selecting those sets of Feynman diagrams corresponding to ladders. In LLA, such
ones are constructed with reggeized gluons in the $t$-channel and bare gluons
on the $s$-channel (the rungs), which are connected by a non-local
gauge-invariant effective vertex. The resultant physical picture is that the
color singlet exchange is associated to a gluon ladder with infinite rungs
\cite{bfkl}. The main result is that the total  cross section for the
exchange process is a power of the center of mass energy, which leads to the
mathematical definition of the BFKL Pomeron as a cut rather than a pole in the
complex angular momentum plane \cite{regge}. 

Such behavior is inconsistent with the requirement of the unitarity bound
\cite{unitarity} and a unitarization procedure has to be performed. The
unitarity constraint states that the total cross section may not grow faster
than $\ln^2 (s)$. Therefore, corrections in order to avoid unitarity violation
present in the amplitude (i.e. total cross section) in the BFKL approach
should be taken into account. In the BFKL approach the violation of unitarity is due
to the fact that the $s$-channel cutted amplitudes contain only a subset of all
the possible intermediate states, namely only gluons in the leading
logarithmic approximation and  gluons plus a $q\bar{q}$ pair in the
next-leading approximation (NLLA). Therefore, we are unable to restore 
unitarity in the BFKL approach even in the NLLA.  In Ref. \cite{ngluon} the
restoration is based on using unitarity and dispersion relations from the
start as a tool to construct higher order amplitudes. The main result of this
approach is the need to take into account contributions with higher number of
reggeized gluons in the $t$-channel, compared to the BFKL amplitude with two
reggeized gluons. In Ref. \cite{multip}, the unitarity problem could be solved
by resumming all multiple BFKL pomeron (at LLA) exchanges in the total cross
section. Although the intense theoretical work at present, the unitarization
problem still remains an open question.

A priori, BFKL is itself asymptotic and we may ask if at finite
energies, i.e. non asymptotic regime, a finite sum of the BFKL series  
could describe the existent data. Recently, Fiore et al. \cite{fiore}
performed a reasonable fit to data on $pp$($p\bar{p})$ total cross section
using this hypothesis. They considered the $n$-rungs ladders diagrams,
with $n=0,1,2$ and $3$, at distinct energy intervals and the parameters are
fitted for each interval. Such procedure introduces a large set of parameters.
An additional fact is that contributions from sub-leading
diagrams in the perturbative expansion are absorved into the parameters. These
features turn the analysis involved when one considers unitarization or
calculation of non-forward observables, as the elastic differential cross
section. 

A well known property of perturbative QCD calculations is that there are
several reasons to believe that the region $t\rightarrow 0$ plays a very special role and
perturbation theory may even not be applicable. Although this fact, in the
recent literature the forward region in hadronic collisions is
treated based on the scale anomaly of QCD, mantaining a perturbative approach
supported by a large scale from the QCD vacuum \cite{scale}, obtaining
consistent results with those ascribed phenomenologically to the soft Pomeron.
In our case, despite the restrictions imposed by the use of a perturbative
description for soft observables, there is sufficient motivation to perform a
deeper analysis on the  BFKL series. In order to make this we should use the
set of diagrams producing contributions $\sim
[\alpha_s\,\ln(s)]^n$,  order by order from perturbation theory, performing a
finite sum of gluon ladders. We notice that when one refers to ladders we have
in mind that they are contructed by reggeized gluons and effective vertices.
The question that remains is how many orders to take into account. The lowest
order two gluons exchange calculation leads to a total cross section constant
on energy. This is a crude approximation to the reality, since experimentally
the cross section has a slow increase with the energy and therefore higher
order contributions are necessary. 

The next contribution to the sum is the one rung gluon ladder. This
calculation provides a  logarithmic growth of the total cross section on
energy, scaled by the typical gluon transverse momentum (in LLA it
is arbitrary). In order to avoid unitarity violation and by simplicity we
truncate our summation at this order, supported by the knowlegde that a
logarithm behavior is in agreement with the experimental results from a
dispersion relation fit \cite{Augier}. In our case the  selected diagrams
cover all energy range, instead of defining a distinct set of diagrams for
sub-intervals of energy as in  Ref. \cite{fiore}. As a result we performed a
successfull fit to the proton-proton(antiproton) total cross sections with
these two contributions. These results motivate to check the non-forward
amplitude in order to obtain the prediction for the elastic differential cross
section, which gives the behavior on the momentum transfer $t$.

In the BFKL framework such analysis is dependent of both the proton impact
factor input and the Lipatov kernel. The BFKL kernel, i.e. Green's function for
the reggeized gluons, is not physical but is under control since it is
calculated from perturbative QCD. For example, the cancelation of the infrared
singilarities in the kernel is known from Ref. \cite{FL}. The amplitude
describing the interaction of the particles (colorless) is the convolution of
the kernel with  the corresponding impact factors and it should be infrared
safe.  In our case, the main uncertainty arises from modeling the proton
impact factor, which presents non-perturbative content. The impact factors
determine the coupling of the Pomeron to the color singlet hadrons and
necessarily vanish when the transverse momentum of any gluon vanishes, which
is required for the cross section to be finite. The infrared singularities
cancelation in the impact factor of colorless particles has been
demonstrated to next-to-leading order in Ref. \cite{FM}. Moreover, the impact factor plays a crucial rule
in the calculation of the non-forward amplitude, in fact determining its
$t$-dependence.

We calculate the proton-proton(antiproton) elastic scattering amplitude at
non zero momentum transfer $t$ taking into account two distict ans\"atze to the
proton impact factor: the Dirac form factor, which has explicit $t$-dependence and
is decoupled in the gluon transverse momenta, as proposed recently by Balitsky 
and Kuchina \cite{balitsky}. The calculation was also performed with an usual
impact factor \cite{askew}, whose shape is determined by quite general
properties and was considered for comparison. The main resulting features are
discussed, having in mind that a more realistic ansatz to the proton impact
factor is still to be found. 

This paper is organized as follows. In the next section we present a short
review of the formulae concerning the two gluons exchange and one rung
contribution to the BFKL approach, presenting the details of the fit to the
proton-proton(antiproton) total cross sections. In the section 3, one presents 
the results to the non forward elastic scattering amplitude with two distinct
impact factor models and their main features are discussed. The elastic
differential cross section is calculated in the small $t$ approximation and
compared with the experimental data at two distinct energy regimes. In the
last section we present our conclusions.

\section{The truncated BFKL series}

By calculating order by order in perturbation theory, summing over the 
leading logarithms of the center of mass energy $s$, one
obtains the BFKL equation, which describes the scattering process by an
infinite rung gluon ladder exchange (see Fig. 1). In this approach,
called leading logarithm approximation, the Pomeron is obtained considering
the color singlet ladder diagrams whose vertical lines are reggeized gluons
coupled to the rungs (bare gluons) through the effective vertices. The
correspondent amplitude is purely imaginary and the coupling constant
$\alpha_s$ is considered frozen in some transverse momentum scale. 

For the elastic  scattering of a hadron, the Mellin transform of the
scattering amplitude is  given by \cite{forshaw}:

\begin{figure}
\centerline{\psfig{file=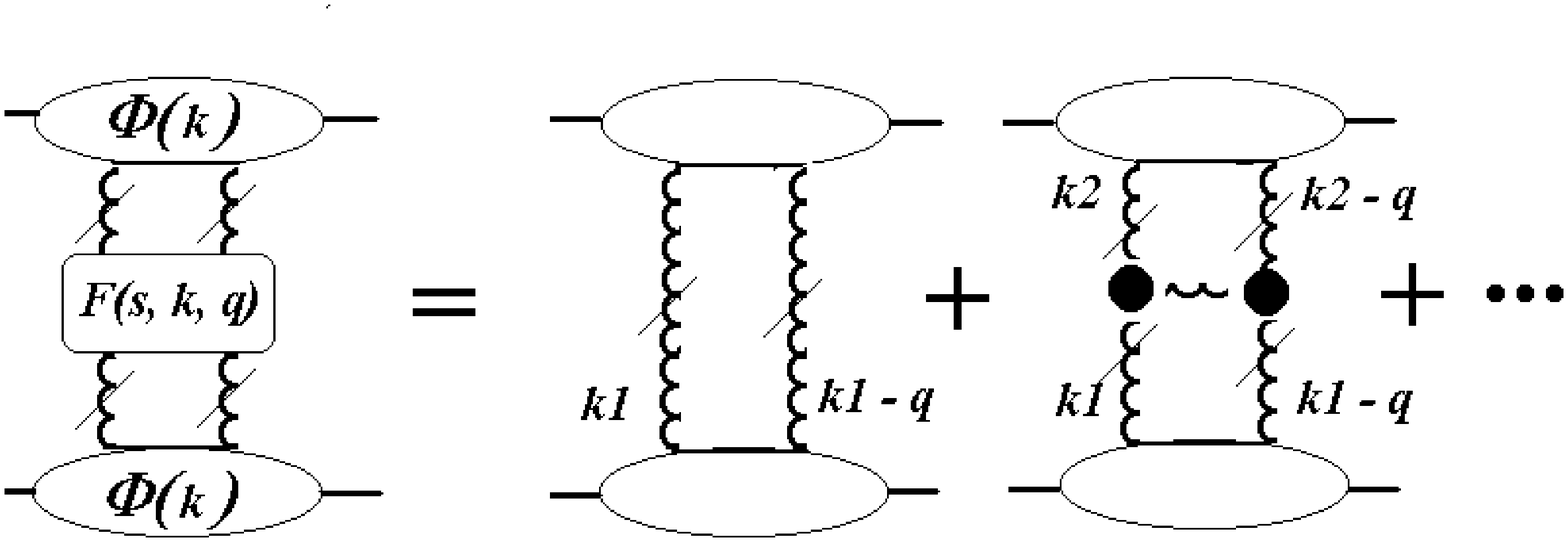,width=130mm}}
\vspace{-2cm}
\caption{The blobs denote the proton(antiproton) structure (Impact
Factors) and the first two orders in perturbation theory are shown. In LLA,
the ladder is constructed with reggeized gluons in the $t$-channel and bare
gluons on the $s$-channel (the rungs), which are connected by a non-local
gauge-invariant effective vertex (the bold blob).}
\end{figure}

\begin{eqnarray} 
{\cal A}(\omega, t) =  \frac{{\cal G}}{(2 \pi)^2} \int d^2 \kk \, d^2 \kg
\,\frac{\Phi(\kk)\Phi(\kg)}{\kg^2(\kk-\qq)^2}\,f(\omega,\kk,\kg,\qq)  \,,  
\label{eq1}
\end{eqnarray}
where the ${\cal G}$ is the color factor for the color
singlet exchange, $\kk$ and $\kg$ are the transverse momenta of the exchanged
gluons in the $t$-channel and $\qq$ is the momentum transfer, with $\qq^2=-t$.
The impact factors describing the interacting particles transition in the
particle-Reggeon (i.e, the reggeized gluons) processes are by definition
factorized from the Mellin transform of the Green's function for the
Reggeon-Reggeon scattering. As a consequence, the energy dependence is
determined by the function $f(\omega,\kk,\kg,\qq)$. This fact  turns evident
once  one defines the transform: 
\begin{eqnarray}
f(\omega)=\int_1^{\infty}\,d\left(\frac{s}{{\bf
k}^2}\right)\,\left(\frac{s}{{\bf k}^2}\right)^{-\omega -1}\,F(s)\,\,.
\label{eq2}
\end{eqnarray} 

In Eq. (\ref{eq2}), a function that is a pure power of $s$ produces a simple
pole on $\omega$; otherwise, as a power of $\ln\,s$ the
transform has a cut sungularity. Therefore the s-dependence of
the amplitude is obtained from the singularity structure of the
transforms.

The function $f(\omega,\kk,\kg ,\qq)$  is the Mellin transform of the BFKL kernel 
$F(s,\kk,\kg,\qq)$, which states the dynamics of the process and is
completely determined in perturbative QCD. The main properties of the LO
kernel are well known \cite{bfkl} and the results arising from the NLO
calculations have yielded intense debate in the literature recently \cite{NLO}.

In the case of $pp$($p\bar{p}$) scattering, the factor $\Phi({\bf k})$ is the
proton impact factor, which in the absence of a perturbative scale has a
non-perturbative feature and provides the Pomeron-proton coupling. This
factor avoids the infrared divergences arising from the
transverse momentum integration. However, it introduces some uncertainty in the
amplitude calculation  since it is not obtained from QCD first principles.  

In the leading order of  perturbation theory we have 
\begin{eqnarray}
f_1(\omega,\kk,\kg,\qq) = \frac{1}{\omega}\,\delta^2(\kk-\kg)\,,
\label{eq3}
\end{eqnarray} and in the next order \begin{eqnarray} f_2(\omega,\kk,\kg,\qq)=
-\frac{\bar{\alpha}_s}{2 \pi}\,\frac{1}{\omega^2} \left[
\frac{\qq^2}{\kk^2(\kg-\qq)^2} -
\frac{1}{2}\frac{1}{(\kk-\kg)^2}\left(1+\frac{\kg^2(\kk-\qq)^2}{\kk^2(\kg-\qq)^2} \right) \right]\,.
\label{eq4}
\end{eqnarray}

For convenience we define $\bar{\alpha}_s=N_c\alpha_s/\pi$, where $N_c$ is the
color number and $\alpha_s$ is the strong coupling constant fixed at
transverse momentum scale.  In order to perform a reliable calculation the
convenient proton impact factor should be introduced. This is not an easy
task, namely these hadronic processes are soft and there is no hard scale
allowing to use perturbation theory.  In fact, we should know in details the
parton wavefunction in the hadron to calculate the impact factors properly.
Since this is not available, several models are proposed in order to calculate
them. This problem is addressed in the next section.

Now we study the results produced when one considers a truncation of the full
BFKL series at the forward (or near forward) region, i.e. $t=0$. The
scattering amplitude, Eq. (\ref{eq1}), can be used to calculate the
$pp(p\bar{p})$ total cross sections. From the Optical Theorem the relation
between the total cross section and the scattering amplitude is~$\sigma_{tot} =
\Im m {\cal A}(s, t = 0)\,/s$, having the lowest order contribution (Born
level) as a constant term in energy, and the next order term  as a logarithm,
scaled by a typical gluon transverse momentum of the process (bearing in
mind it is arbitrary). When considering zero momentum transfer there is no need
to deal with both a specific form for the impact factor and the transverse
momentum integration. This allows to consider $s$-independent factors in each
term as free parameters and to obtain them from data.

\begin{figure}
\centerline{\psfig{file=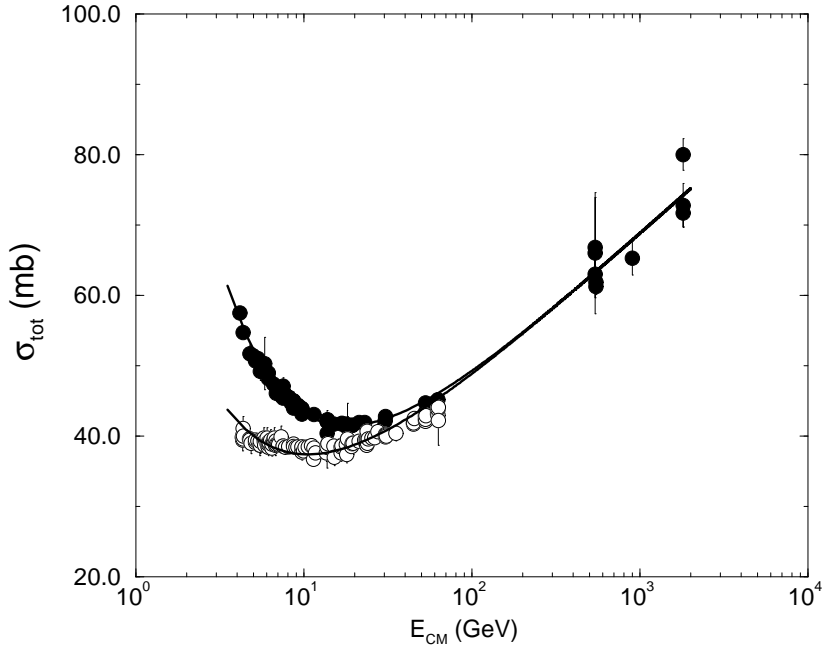,width=120mm}} 
\caption{Result of the $pp(p\bar{p})$ total cross sections \cite{data}.
The errors are summed into quadrature.} 
\end{figure}

We select the set of data on proton-proton(antiproton) total cross section
\cite{data}, considering points with $\sqrt{s} > 4\,GeV$ to avoid very low
energy data, and  choose the typical transverse momentum as ${\bf k^2}=s_0=1\,
GeV^2$, in such a way that the factors are in $(mb)$. The correct description
at low energy requires the reggeon contribution, which is parameterized from
Regge theory. Our expression to the total cross section is then,

\begin{eqnarray} 
\sigma_{tot}^{pp(p\bar{p})}= C_{R}\,(s/s_0)^{\alpha_{R}(0)-1} +
C_{Born} + C_{NO}\,\ln (s/s_0)\,. 
\end{eqnarray}
The reggeon intercept at
zero momentum transfer is $\alpha_{R}(0)$ and the factor $C_{R}$ is distinct
to $pp$ and $p\bar{p}$, as a consequence of the different reggeon coupling to
particle  and antiparticle. Consistent with the usual Donnachie-Landshoff
fit \cite{dola} for $p\bar{p}$, the reggeon contribution
is described effectivelly by $\alpha_{R}=0.5475$. Hence we fix the constants
$C_{Born}$ and $C_{NO}$ from data on $p\bar{p}$, imposing the same
contribution for both proton-proton and proton-antiproton. This procedure is
reasonable due to the higher energies reached on $p\bar{p}$ collision, where
the Pomeron dominates. On the other hand, $pp$ data are predominantly at low
energy, which is not strongly sensitive to the Pomeron model, thus dominated
by the  reggeonic contribution. In the $pp$ case there is need  of a more
refined parameterization for the reggeonic piece \cite{dola}, therefore we
consider the intercept as a free parameter for this process. A successful
description of data  is obtained for the whole range of energy. The result is
shown in the Fig. (2), and the parameters are presented in Tab. (1).

\begin{table}[h]
\begin{center}
\begin{tabular} {||l|l|l|l|l||}
\hline
\hline
 Process & $\;\;C_R$ & $\;\;\;\;\;\alpha_R(0)$ & $C_{Born}$ & $C_{NO}$
\\ \hline $p\bar{p}$& 141.51 & 0.5475 (fixed) & 4.16 & 4.66 \\
\hline
$pp$ & 78.15 & 0.589 (free) & 4.16  & 4.66  \\
\hline
\hline
\end{tabular}
\end{center}
\caption{The parameters for $p\bar{p}$($pp$) cross sections ($
\chi^2=1.19$).}   \end{table}

Our result is similar to that of Fiore et al. \cite{fiore}, with the
 advantage to use a smaller set of parameters (see Tab. 1), and no
additional soft Pomeron is used in our analysis. We use only two orders in
perturbation theory (up to the one rung ladder) while the authors in Ref.
\cite{fiore} use up to the fourth order. We describe the $p\bar{p}$ total cross
section with only three free parameters (five for $pp-p\bar{p}$ simultaneous
fit), instead of 12 or 16 (considering up to two rungs or three rungs ladder,
respectively) from Ref. \cite{fiore}. An important additional advantage is that
the total cross section obtained is consistent with the unitarity constraint,
avoiding eikonalization procedures. 

The hypothesis of considering two orders from the BFKL series, namely 
$\sigma_{tot}\sim \ln (s)$, is phenomenologically corroborated by the well
known dispersion relation fit \cite{Augier}. This fit is based on measurements
of $\sigma_{tot}$ and $\rho$-parameter in energies $5\,GeV \leq \sqrt{s}\leq
546\,\,GeV$ and the high energy dependence was described by $\sim [\log
(s/s_0)]^\gamma$, with $\gamma=2.2 \pm 0.3$. A simple logarithmic behavior,
$\gamma=1$, is favoured by the results of the experimental group E710/E811 \cite{E710} at
$\sqrt{s}=1800\,\,GeV$ and supported by the very high energy cosmic ray data \cite{cosmray}. 
As a final remark, at the LHC energy ($\sqrt{s}=14\,\,TeV$) the extrapolation
of our results will give $\sigma_{tot}=93.22\,\,mb$.

In the next section we calculate the elastic amplitude at non zero momentum
transfer using two distinct models for the proton impact factor, discussing
its main properties and obtaining a description of the existent data in the
small-$t$ approximation.

\section{The non forward scattering amplitude}

Now we perform an analysis of the elastic scattering amplitude at non zero
momentum transfer $t=-\qq^2$. In order to calculate this amplitude, information
about the coupling between the proton and the $t$-channel gluons in the
ladder is required. Namely, we should introduce a reliable proton impact
factor. 

In the calculation of the hadron-hadron scattering amplitude the basic diagram
is the quark-quark elastic scattering, which are taken on shell. This fact does
not correspond to reality since the Pomeron couples to the hadron whose
constituent partons are slightly off-shell. For the quark-quark case,  although
$f(\omega,\kk,\kg,\qq)$ does not contain any infra-red singularities, the
amplitude nevertheless diverges due to the remaining integrals over $\kk$ and
$\kg$ which develop infra-red singularities when  $\kk$ and $\kg$ (or 
$(\kk-\qq)$, $(\kg-\qq)$)   go to zero.  In principle, when we convolute the
bare amplitude with the impact factors it should be infra-red safe. The
next task is to model the impact factor since it cannot be
calculated from first principles due to the unknowledge on the wavefunction of
the hadronic constituent partons. 

Here are analized two distinct models for the impact factor and its
consequences for the elastic amplitude and the differential cross section.

\subsection{Dirac form factor:}
Balitsky and Kuchina proposed recently \cite{balitsky} that at large momentum
transfer the coupling of the BFKL Pomeron to the nucleon is essentially equal
to the Dirac form factor of the nucleon. Their basic idea is that in the
lowest order in perturbation theory there is no difference between the
diagrams for the nucleon impact factor and similar diagrams with two gluons
replaced by two photons, in such a way that the amplitudes  can be calculated
without any model assumption.

This impact factor is  decoupled in the transverse
momentum integration and presents an explicit dependence on $t$, being
similar to the usual Pomeron-proton coupling used in Regge phenomenology. The
expression is

\begin{eqnarray}
\Phi_p({\bf k},\qq)= F_1^{p+n}(t) =
\frac{1}{1+\left(\frac{|t|}{0.71\,GeV^2}\right)^2}\frac{4m_p^2+0.88|t|}
{4m_p^2+|t|}\,. \end{eqnarray}

The choice for this proton impact factor is useful when one analyzes near
forward observables, for instance the elastic differential cross section.
However,  it does not play the role of a regulator of infrared divergences
arising from the calculations at proton-proton(antiproton) process, because
clearly it does not vanish when the gluon transverse momenta goes to zero. In
electron-proton process the situation is different  since the photon impact
factor supplies that condition \cite{balitsky}. 

Then the next step is to perform the gluon transverse momenta integrations. In
fact, such integrals are infrared divergent and should be regularized. An
usual way out is to introduce an  infrared cut-off $\lambda^2$ (for instance, 
Ref. \cite{regula}), temporally defining a small gluon mass, avoiding
problems at the infrared region. This procedure is quite similar as to take
into account a non-perturbative massive gluon propagator (i.e., see Ref.
\cite{halzen}).  

The lowest order (order $\alpha_s^2$) contribution, i.e. the Pomeron at the
Born level, gives the following result:
\begin{eqnarray} {\cal A}^{(1)}(s,t \,; \lambda^2)= \frac{{\cal
G}^{\prime}}{(2\pi)^4}\,s \, \int d^2 {\bf k}\,\frac{\Phi_p^2({\bf k})}{{\bf
k}^2({\bf k}- \qq)^2}=\frac{{\cal G}^{\prime}}{(2\pi)^4}\,s\, [F_1^{p+n}(t)]^2
\, \frac{\pi}{(|t|-\lambda^2)}\,\ln \left(\frac{\lambda^2}{|t|}\right) \,.
\nonumber \end{eqnarray}

Here we notice that there is an implicit dependence on $\lambda^2$ in the
above equation. The one rung gluon ladder has two components (order
$\alpha_s^3$), given by the following expression: \begin{eqnarray} {\cal
A}^{(2)}(s,t \,; \lambda^2)= \frac{{\cal
G}^{\prime}}{(2\pi)^4}\,s\,[F_1^{p+n}(t)]^2 \ln\left(\frac{s}{{\bf
k^2}}\right)\,(I_1 + I_2)\,, \end{eqnarray} with $I_1$ corresponding to the
one rung gluon ladder and $I_2$ correspondent to the three gluons exchange
graphs, whose order is also $\ln(s/{\bf k^2})$. Such structure is due to the
fact that in the color singlet calculation there is no cancellation between
graphs and one can not  obtain an expression for the two-loop level which is
proportional to the one loop amplitude \cite{forshaw}. We define $I_2$ through
symmetry on the integration variables $\kk$ and $\kg$ (see equations
(\ref{eq1}-\ref{eq4})) and the factor ${\cal G}^{\prime}$ collects the
correspondent color factors and the remaining constants. The explicit
calculation of those integrals, yields

\begin{eqnarray}
I_1= -\qq^2\int d^2 \kk \frac{1}{\kk^2(\kk-\qq)^2} \int d^2 \kg
\frac{1}{\kg^2(\kg-\qq)^2} = -\pi^2 \frac{|t|}{(|t|-\lambda^2)^2}\,\ln^2
\left(\frac{\lambda^2}{|t|} \right)\,, \nonumber \end{eqnarray}

\begin{eqnarray}
I_2= \int d^2 \kk  \int d^2 \kg \frac{1}{\kk^2(\kk-\kg)^2 (\kg-\qq)^2}=
\frac{1}{2}\frac{\pi^2\,\ln(\lambda^2)}{(|t|-\lambda^2)}
\ln\left(\frac{\lambda^2}{|t|}\right)\left(
1-\frac{\ln(|t|)}{\ln(\lambda^2)}\right)\,. \nonumber \end{eqnarray}

Some comments about the amplitude above are in order. The scale for the factor
$\lambda^2$ should be at non-perturbative regime, i.e. $\lesssim 1\,\,GeV^2$.
In the Fig. (3) we show a comparison between the predicted differential cross
section using the Balitsky-Kuchina impact factor and the experimental results
at 1800 GeV. An analysis is performed for two distinct values of the cut-off
$\lambda^2$. The prediction presents a deviation of the usual exponential
parameterization from Regge phenomenology and a remarkable difference appears
at larger $t$ values. In addition, the impact factor $\Phi_p({\bf k}, \qq)$ 
above does not satisfy the condition $\Phi({\bf k}=0,\qq)=\Phi({\bf
k}=\qq,\qq)=0$, required for the corresponding cross section to be finite
\cite{FM}, giving rise to the singularity at $t=0$ for the calculated
amplitude.

\begin{figure}
\centerline{\psfig{file=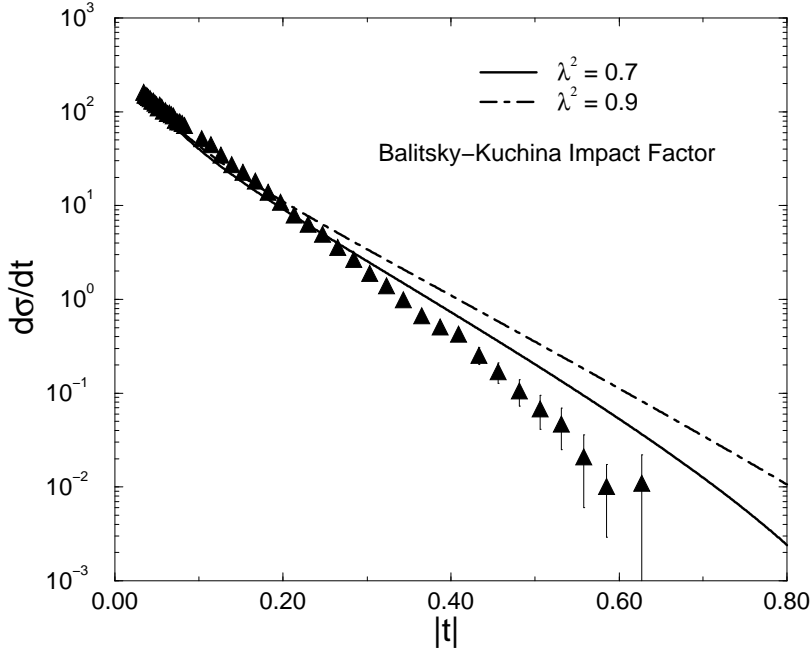,width=120mm}} 
\caption{ The predicted differential cross
section using the Balitsky-Kuchina impact factor [11] and the experimental
results at 1800 GeV, for two distinct values of the cut-off
$\lambda^2$.} 
\end{figure}

Moreover, an interesting aspect is the behavior of the amplitude at the forward
limit $t=0$, where it becames very large. This limit is a well known property
of perturbative QCD calculations and there are several reasons to believe that
the point $t=0$ plays a very special role, such that perturbation theory may
even not be applicable. Concerning the forward region, for the full BFKL
series there is still the diffusion on tranverse momenta, i.e. on $\ln {\bf
k}^2$, which extends into both the ultraviolet and the infrared regions
\cite{infrared}. Nevertheless, the momentum scale $t$ supplies  the control
condition.

However, we suppose that a smooth transition from a finite $t$ down to
$t=0$ is possible and that the truncated BFKL series gives the correct behavior
on energy for the forward observables. Later we make use of this hypothesis to
get a parameterization to the logarithmic slope $B(s)$ and calculate the
differential elastic cross section.  

\subsection{Usual non-perturbative ansatz:}

Using quite general properties of the impact factors, namely they vanish as
the $t$-channel gluons transverse momenta go to zero, one can guess their
behavior which is determined by the large scale nucleon dynamics.
Such study has been  performed at Ref. \cite{askew}, where the solutions of
the Lipatov equation are examined critically and determined their importance on
the structure function description using physically motivated modifications
for small ${\bf k}^2$. Namely, it was performed a detailed parametrization of
the infrared region which satisfies the gauge invariance constraints when
${\bf k}^2 \rightarrow 0$. We use this result to study its role in our
calculation for the elastic amplitude. The impact factor is written now as,  

\begin{eqnarray}
\Phi_p({\bf k})= \frac{{\bf k}^2}{{\bf k}^2 + \mu^2}\,,
\end{eqnarray}
where $\mu^2$ is a scale which is typical of the non-perturbative dynamics and
is related to the radius of the gluonic form factor of the proton. Considering
it as the scale of the hadronic electromagnetic form factor, then ${\bf k}^2
\simeq 0.5 \,\,GeV^2$ instead of estimates from QCD sum rules giving ${\bf k}^2
\simeq 1-2\,\,GeV^2$ \cite{askew}.

 As a consequence of this choice for the impact factor,  the
momentum transfer behavior is completely determined by the kernel, since we
consider $\qq \neq 0$. The amplitude now reads
\begin{eqnarray} {\cal
A}(s,t)& = & \frac{{\cal G}}{(2\pi)^4}\,s\,\pi\,I_1(t,\mu^2) + \frac{{\cal G}}{(2\pi)^4}\,s\,\pi \ln\left(\frac{s}{{\bf
k}^2}\right)\, \left[ I_1^2(t,\mu^2) + I_2(t,\mu^2) \right] \,.
\label{eq12}
\end{eqnarray}
 \noindent where

\begin{eqnarray}
I_1(t,\mu^2)&=&\frac{1}{(|t|-\mu^2)} 
+ \frac{|t|} {(|t|-\mu^2)^2} \ln \left(\frac{\mu^2}{|t|}\right)\,\,,
\nonumber \\ 
I_2(t,\mu^2)&=&\frac{\ln (\mu^2)}{(|t|-
\mu^2)} + \frac{\ln(\mu^2)|t|}{(|t|-\mu^2)^2}\ln
\left(\frac{\mu^2}{|t|}\right)\,\,.\nonumber
\end{eqnarray}

In the Fig. (4) we present the prediction to the
differential cross section at 1800 GeV, using two distinct values for the
parameter $\mu ^2$. Again a deviation from the exponential parametrization
based on  Regge phenomenology is present, mainly at larger $t$. We notice
that a rather crude approximation for the Pomeron-proton coupling can be
improve this result (for instance, an exponential parameterization). We observe
again a  divergent behavior at $t=0$, as a consequence of the impact factor
which does not satisfy the condition $\Phi({\bf k}=0,\qq)=\Phi({\bf
k}=\qq,\qq)=0$.

\begin{figure}
\centerline{\psfig{file=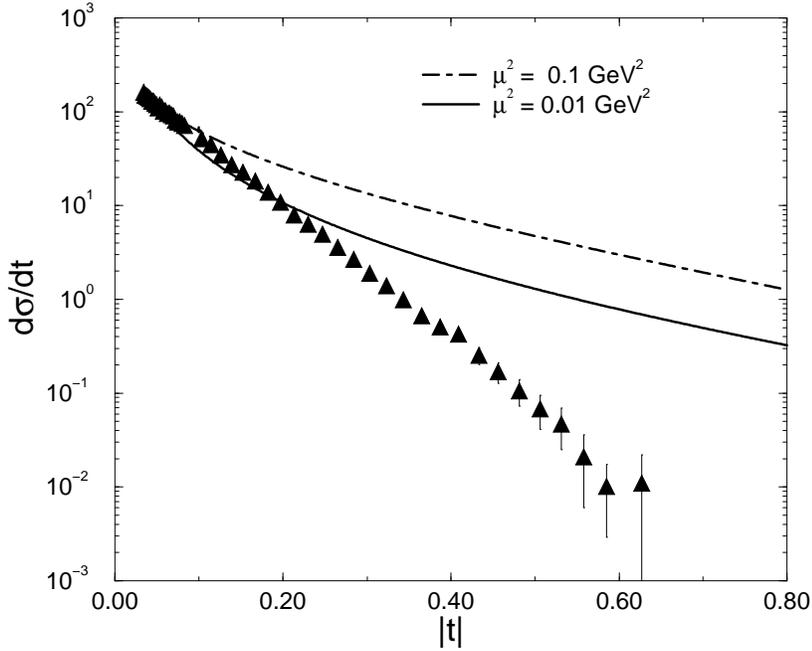,width=120mm}} 
\caption{ The prediction to the
differential cross section using the usual impact factor [12] and the
experimental results at 1800 GeV [15], using two distinct values for the
parameter $\mu^2$.}  \end{figure}

 Despite obtaining an analytic expression to the elastic scattering
amplitude, i.e. differential cross section, a direct comparison with the whole
 experimental data is known not to be reliable. To perform a more
phenomenological analysis we notice that one can parametrize the elastic slope
claiming that the forward amplitude is finite in $t=0$ and the dependence on
energy is correctly described by the truncated BFKL series. Such hypothesis is
supported by the factorization on  energy and momentum transfer present in
the amplitudes. In addition, data on differential cross section at low $t$ are
parameterized in the form $d\sigma/dt=A\,e^{B\,t}$, where $B$ is the forward
slope \cite{matthie}. Therefore, we can obtain an expression for the
differential cross section at small $t$, using our previous results. 

The usual relation to
describe the cross section is:
 \begin{eqnarray} \frac{d\sigma^{el}}{dt} & = &
\frac{d\sigma}{dt}|_{t=0}\,e^{B(s,\,t=0)\,t}=\frac{\sigma_{tot}^2}{16
\pi}\,e^{B_{el}(s)\,t}\,,\\ B(s,t=0) & = &
\frac{d}{dt}\left[log\,\frac{d\sigma}{dt}\right]\,. \end{eqnarray} 

In the
Regge framework the $B$ slope is obtained from the powerlike behavior of the
scattering amplitude, dependent of the effective slope of the Pomeron
trajectory $\alpha^{\prime}_{P}$, namely $B_{el}^{Regge}(s)= 4\,b_0 +
2\,\alpha^{\prime}_{P}\,\ln(s)$. The $b_0$ comes from the exponential
parameterization for the slope of the proton-proton-Pomeron vertex. In our case
we should calculate the slope from the non forward elastic scattering
amplitudes ${\cal A}^{Ladder}(s,t)$ obtained above. For the amplitude obtained
employing  he Balitsky and Kuchina impact factor one obtains the following
slope  \begin{eqnarray} B(s)=\frac{4}{F_1^{p+n}(t)}\,\frac{d
F_1^{p+n}(t)}{dt}|_{t=0} + \frac{2}{{\cal A}^{Ladder}}\frac{d\,{\cal
A}^{Ladder}}{dt}\,\,|_{t=0} \,,  \end{eqnarray} where the first term  does not
contribute effectively at $t=0$ and we are left only with the second term.
From simple inspection of the amplitude obtained with the usual impact factor
(see Eq. \label{eq12}) we also verify that one gets a similar expression to the
correspondent slope. 

Considering the  specific form for the $t$-derivative of the amplitudes, their
asymptotic values at $t=0$ depend only on the energy. In fact, they take the
form $d{\cal A}/dt=R_1\,s+ R_2\,s \ln(s/s_0)$, where $R_1$ and $R_2$ are
$s$-independent parameters. For our case, the amplitude is purely imaginary,
then $|{\cal A}(s,t=0) |= s\,\sigma_{tot}$ and $d\sigma/dt
\,|_{t=0}=\sigma_{tot}^2/16\pi$. Putting all together, the corresponding slope
and the elastic differential cross section are \begin{eqnarray}
B(s) & = & \frac{2}{\sigma_{tot}}\left[R_1 + R_2\,\ln(s/s_0)\right]\,\,,\\
\frac{d\sigma}{dt} & = & \frac{\sigma_{tot}^2}{16\pi}\,e^{B(s)\,t}\,,
\end{eqnarray} where again $s_0=1\,GeV^2$. 

In order to obtain the parameters
$R_1$ and $R_2$, we use the slope experimental values for both low (CERN-ISR)
and high energy (CERN-SPS,Tevatron) points from $p\bar{p}$ reaction
($23\,<\sqrt{s}<\,1800\,\,GeV$) \cite{data}. The total cross section is given
by Eq. (\ref{eq5}). Our result is shown in the Fig. (5), and the parameters
are presented in Tab. (2). For completeness we include the reggeon contribution
since we also deal with low energy data, requiring one  additional parameter
($b_R$) coming from the parameterization to the proton-proton-Reggeon vertex. 

\begin{figure}[t] 
\centerline{\psfig{file=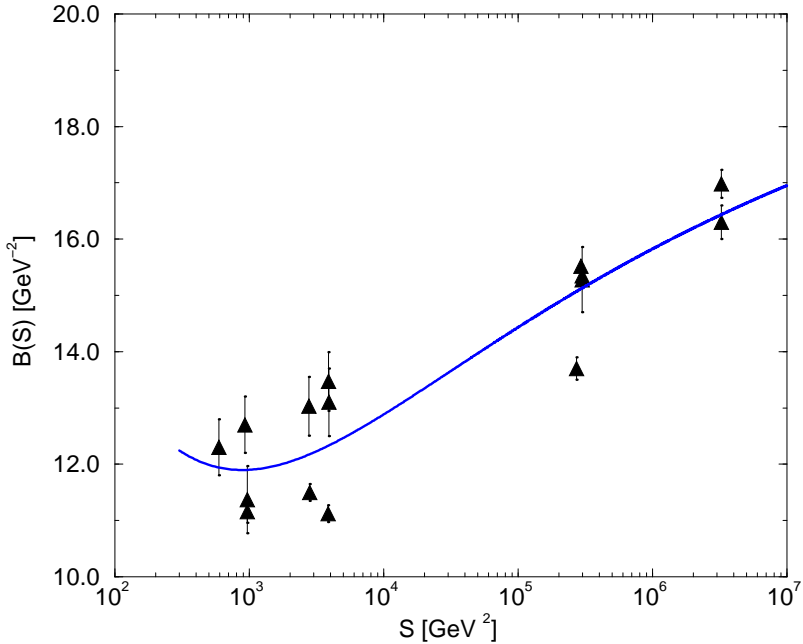,width=120mm}} 
\caption{The result for the slope $B(s)$, using both low and high energy data
points on $p\bar{p}$ reaction \cite{data}.}  
\end{figure}

\begin{table}[h]
\begin{center}
\begin{tabular} {||l|l|l|l||}
\hline
\hline
Process & $b_R$ ($mb$) & $R_1$ ($mb^2$)& $R_2$ ($mb^2$) \\
\hline
$p\bar{p}$& $\;\;4.62$  & $\;\;-99.7$  & $\;\;22.39$ \\
\hline
\hline
\end{tabular}
\end{center}
\caption{The parameters for the $p\bar{p}$ forward slope $B_{el}(s)$ ($
\chi^2=0.71018$).} \label{turn} 
\end{table}

Having the slope obtained from data, the elastic differential cross section is
straighforwardly determined and a successful comparison with its experimental
measurements at $\sqrt{s}=53\,GeV$ and $\sqrt{s}=1800\,GeV$ is shown in the
 Fig. (6).

A reliable description of both total and differential cross
sections is obtained, allowing the study of the role played by the impact
factors in the calculations, for instance the factorizable feature of the
Balitsky and Kuchina impact factor. 

\begin{figure}[t]
\begin{tabular}{cc}
\psfig{file=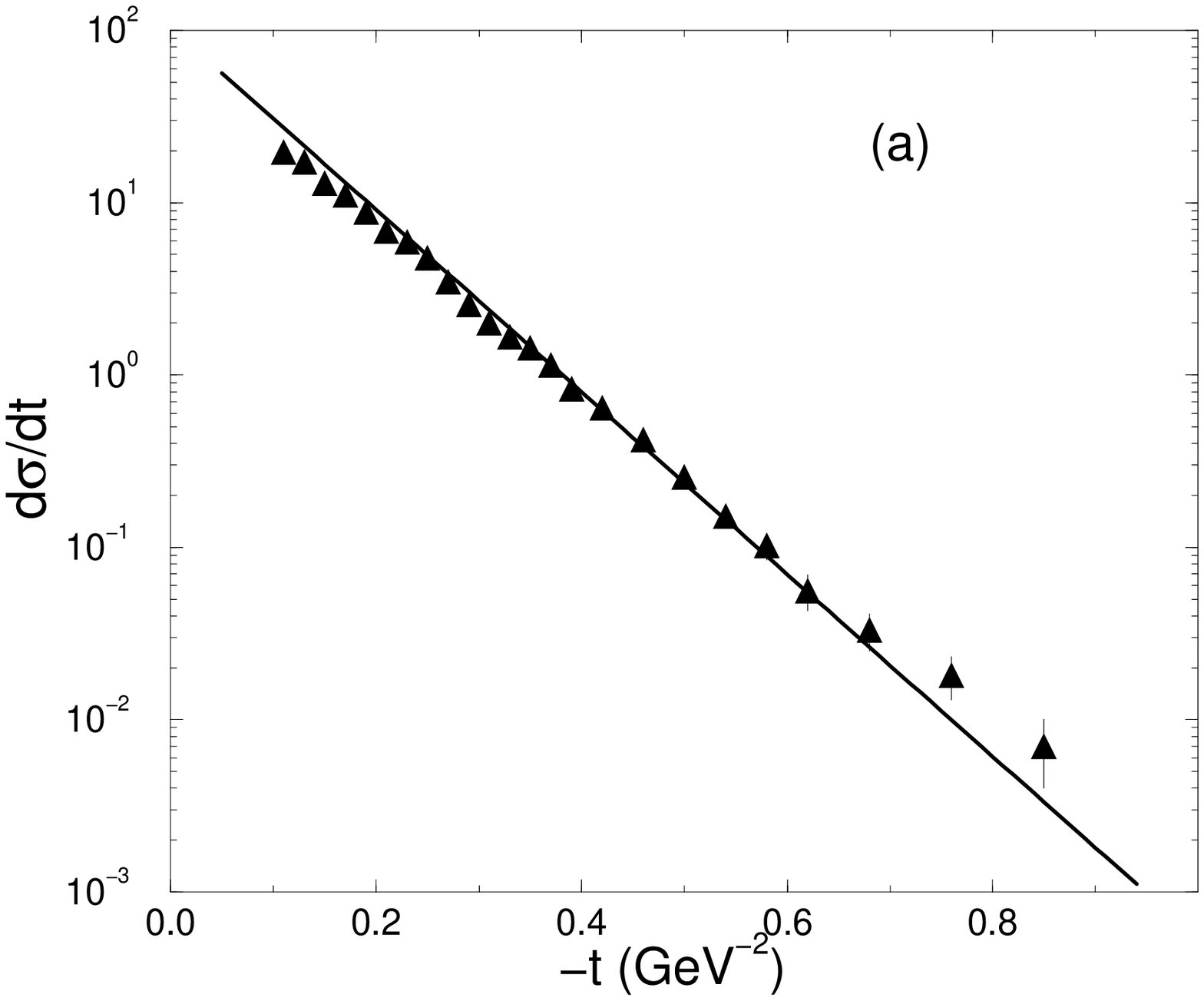,width=77mm}  & \psfig{file=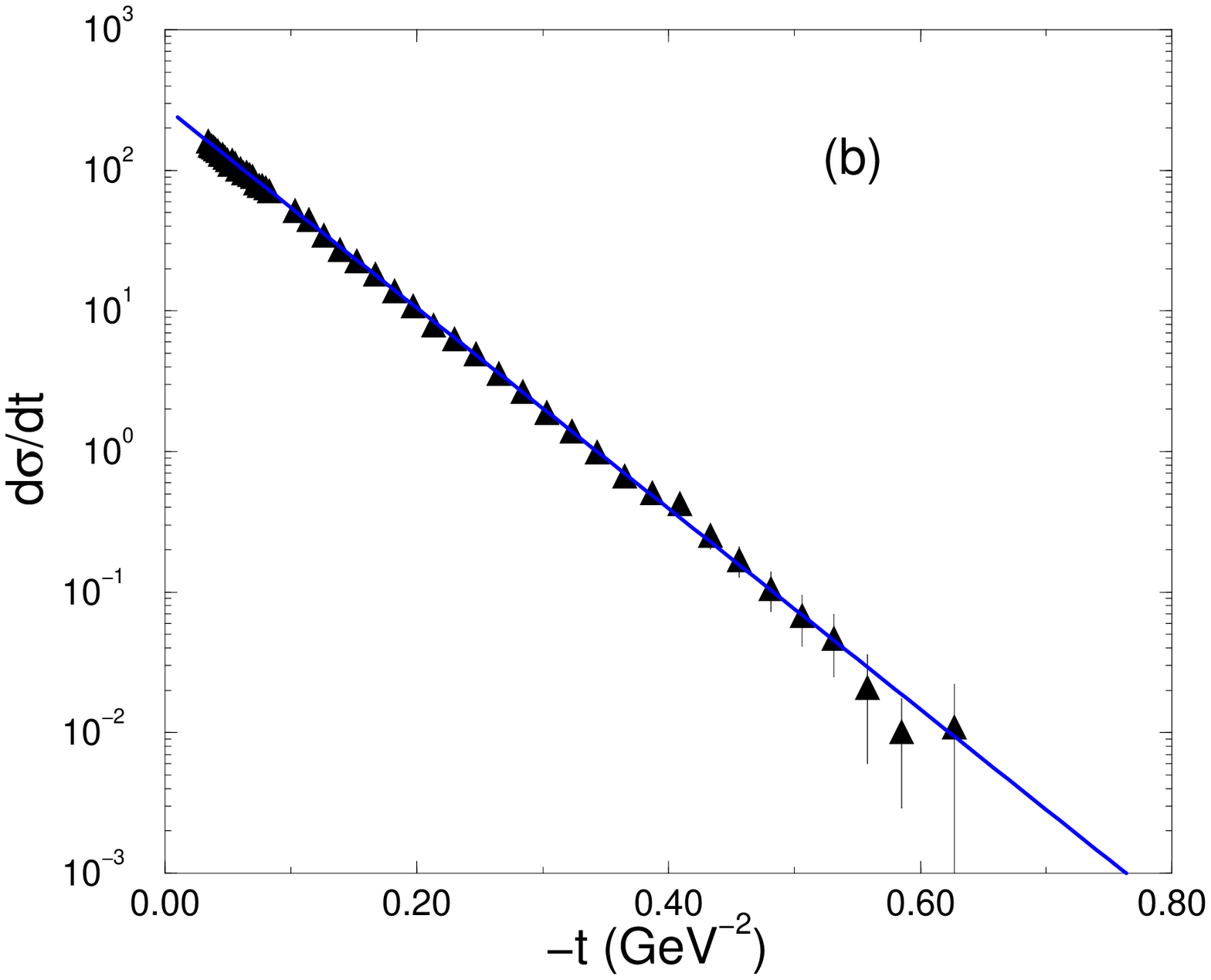,width=77mm} 
\end{tabular}
\caption{The result for the elastic differential cross section at
(a) $\sqrt{s}= 53\,\,GeV$ and (b) $\sqrt{s}= 1800\,\,GeV$ [15].}
\end{figure}

It is well-known that the large $t$ data are dominated by
the perturbative contribution, as verified by Donnachie-Landshoff in the
calculation of three gluons exchange for the $pp$($p\bar{p})$ reactions
\cite{threegluon}. A further analysis will require  the complete
elastic amplitude rather than the small $t$ approximation, i.e., to describe
the large $t$ region and extend the model to a wider interval in the momentum
transfer.  The $pp$ reaction presents the typical dips at momentum transfer of
order 1, 2 $GeV^2$ \cite{threegluon}, which is not included in the small $t$
approximation.  The usual procedure to solve this problem is by eikonalizing
the Born amplitude, whose physical picture is the multiple elastic scattering
of the Pomeron exchange \cite{eikonal}. In the present case, the Born amplitude
does not violate the unitarity constraint and such procedure seems not to be 
necessary. However, the dips structure  can be present in the amplitude, i.e. 
the whole $t$ domain can be described taking a suitable choice
of the impact factor.

\section{Conclusions}
We study in detail the contribution of a  truncated BFKL series to the hadronic process,
specifically the proton-proton(antiproton) collisions, considering two orders
in perturbation theory corresponding to the two reggeized gluons exchange and
the one rung gluon ladder (considering the effective vertex). Despite the
restrictions imposed by the use of a perturbative approach for soft
observables, a good description of the total cross sections was obtained
motivating an analysis of the elastic differential cross section. Although the
QCD perturbation theory is in principle not reliable at the forward direction
($t=0$), nevertheless we suppose that perturbation theory gives the behavior
on energy even in this region.  The next step is to consider $t$ different
from zero, where the momentum transfer furnishes a scale to perform suitable
calculations.  In order to proceed this, we calculate the non forward
amplitude introducing two distinct ans\"atze for the proton impact factor,
namely a factorizable $t$-dependent proposed recently by Balitsky and Kuchina
and the usual non-perturbative impact factor. In order to describe data we
used  a small momentum transfer approximation and obtained an expression to
the elastic slope $B_{el}(s)$, determining the correspondent parameters. The
elastic differential cross section is obtained straightforwardly, describing
with good agreement the experimental data at both low and high energy values.

\section*{Acknowledgments}

This work was partially financed by CNPq and by PRONEX (Programa de Apoio a
N\'ucleos de Excel\^encia), BRAZIL.

\end{document}